\documentclass[aps,prl,twocolumn, amsmath,amssymb,superscriptaddress,nofootinbib]{revtex4-1}
\usepackage[utf8]{inputenc}
\usepackage{graphicx}
\usepackage{xcolor}
\usepackage{amsmath}
\usepackage{ulem}
\usepackage{dcolumn}
\usepackage{bm}
\expandafter\ifx\csname package@font\endcsname\relax\else
\expandafter\expandafter
\expandafter\usepackage
\expandafter\expandafter
\expandafter{\csname package@font\endcsname}
\fi
\begin{document}
\title{Stochastic Inference of Surface-Induced Effects using Brownian Motion}
\author{Maxime Lavaud}
\affiliation{Univ. Bordeaux, CNRS, LOMA, UMR 5798, F-33405 Talence, France.}
\author{Thomas Salez}
\email{thomas.salez@u-bordeaux.fr}
\affiliation{Univ. Bordeaux, CNRS, LOMA, UMR 5798, F-33405 Talence, France.}
\affiliation{Global Station for Soft Matter, Global Institution for Collaborative Research and Education, Hokkaido University, Sapporo, Japan.}
\author{Yann Louyer}
\affiliation{Univ. Bordeaux, CNRS, LOMA, UMR 5798, F-33405 Talence, France.}
\author{Yacine Amarouchene}
\email{yacine.amarouchene@u-bordeaux.fr}
\affiliation{Univ. Bordeaux, CNRS, LOMA, UMR 5798, F-33405 Talence, France.}
\date{\today}
\begin{abstract}
Brownian motion in confinement and at interfaces is a canonical situation, encountered from fundamental biophysics to nanoscale engineering. Using the Lorenz-Mie framework, we optically record the thermally-induced tridimensional trajectories of individual microparticles, within salty aqueous solutions, in the vicinity of a rigid wall, and in the presence of surface charges. We construct the time-dependent position and displacement probability density functions, and study the non-Gaussian character of the latter which is a direct signature of the hindered mobility near the wall. Based on these distributions, we implement a novel, robust and self-calibrated multifitting method, allowing for the thermal-noise-limited inference of diffusion coefficients spatially-resolved at the nanoscale, equilibrium potentials, and forces at the femtoNewton resolution.
\end{abstract}
\maketitle

Brownian motion is a central paradigm in modern science. It has implications in fundamental physics, biology, and even finance, to name a few. By understanding that the apparent erratic motion of colloids is a direct consequence of the thermal motion of surrounding fluid molecules, pioneers like Einstein and Perrin provided decisive evidence for the existence of atoms~\cite{einstein_uber_1905,perrin_les_2014}.
Specifically, free Brownian motion in the bulk is characterized by a typical spatial extent evolving as the square root of time, as well as Gaussian displacements. 

At a time of miniaturization and interfacial science, and moving beyond the idealized bulk picture, it is relevant to  consider the added roles of boundaries to the above context. Indeed, Brownian motion at interfaces and in confinement is a widespread practical situation in microbiology and nanofluidics. In such a case, surface effects become dominant and alter drastically the Brownian statistics, with key implications towards: i) the understanding and smart control of the interfacial dynamics of microscale entities; and ii) high-resolution measurements of surface forces at equilibrium. Interestingly, a confined colloid will exhibit non-Gaussian statistics in displacements, due to the presence of multiplicative noises induced by the hindered mobility near the wall~\cite{Felderhof2005,wang_anomalous_2009,chechkin_Brownian_2017}. Besides, the particle can be subjected to electrostatic or Van der Waals forces~\cite{Bouzigues2008} exerted by the interface, and might experience slippage too~\cite{Joly2006,Mo2017}. Considering the two-body problem, the nearby boundary can also induce some effective interaction~\cite{Dufresne2000}. Previous studies have designed novel methods to measure the diffusion coefficient of confined colloids~\cite{faucheux_confined_1994,Dufresne2001,Eral2010,Sharma2010,Mo2015,matse_test_2017}, or to infer surface forces~\cite{prieve_measurement_1999,Banerjee2005,Sainis2007,volpe_influence_2010,Wang2011,li_subfemtonewton_2019}. However, such a statistical inference is still an experimental challenge, and a precise calibration-free method taking simultaneously into account the whole ensemble of relevant properties, over broad spatial and time ranges, is currently lacking.

\begin{figure}[t!]
\centering
\includegraphics[scale=1]{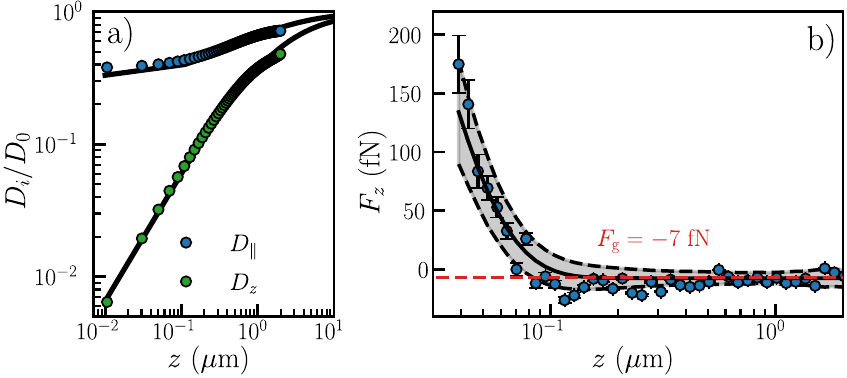}
\caption{a) Measured local short-term diffusion coefficients $D_i$ of the microparticle, normalized by the bulk value $D_0$, as functions of the distance $z$ to the wall (see Fig.~\ref{Fig2}c)), along both a transverse direction $x$ or $y$ ($D_i=D_\parallel=D_x=D_y$, blue) and the normal direction $z$ ($D_i=D_z$, green) to the wall. The solid lines are the theoretical predictions, $D_{\parallel}(z)=D_0\eta/\eta_{\parallel}(z)$ and $D_z(z)=D_0\eta/\eta_z(z)$, using the local effective viscosities $\eta_{\parallel}(z)$ and $\eta_z(z)$ of Eqs.~(\ref{etapara})~and~(\ref{etaperp}), respectively. b) Total normal conservative force $F_z$ exerted on the particle as a function of the distance $z$ to the wall, reconstructed from Eq.~(\ref{stokes}), using Eq.~(\ref{etaperp}). The solid line corresponds to Eq.~(\ref{Eq:Force}), with $B=4.8$, $\ell_{\mathrm{D}}=21\,\mathrm{nm}$ and $\ell_{\mathrm{B}}=530\,\mathrm{nm}$. The black dashed lines and grey area indicate the amplitude of the thermal noise computed from Eq.~(\ref{tnl}). The horizontal red dashed line indicates the buoyant weight $F_{\textrm{g}}=-7$~fN of the particle.}
\label{Fig1}
\end{figure}
In this Letter, we aim at filling the previously-identified gap by implementing a novel method of statistical inference on a set of trajectories of individual microparticles recorded by holographic microscopy. The buoyant particles are free to evolve within salty aqueous solutions, near a rigid substrate, and in the presence of surface charges. We primarily reconstruct the equilibrium probability distribution function of the position, as well as the time-resolved probability distribution functions of the displacements in directions transverse and normal to the wall, including in particular the mean-squared displacements. Special attention is dedicated to the non-Gaussian statistics, for time scales broadly ranging  from tens of milliseconds to several tens of minutes. Furthermore, we implement the advanced inference method recently proposed~\cite{frishman_learning_2020}. Besides, an optimization scheme is used in order to determine precisely all the free physical parameters and the actual distance to the wall, at once. All together, this procedure leads to the robust calibration-free inference of the two central quantities of the problem: i) the space-dependent short-term diffusion coefficients, with a nanoscale spatial resolution; and ii) the total force experienced by the particle, at the thermal-noise limited femtoNewton resolution. These main results are summarized in Fig.~\ref{Fig1}, the goal of the Letter being the detailed obtention of which.

\begin{figure}[t!]
\centering
\includegraphics[scale=1]{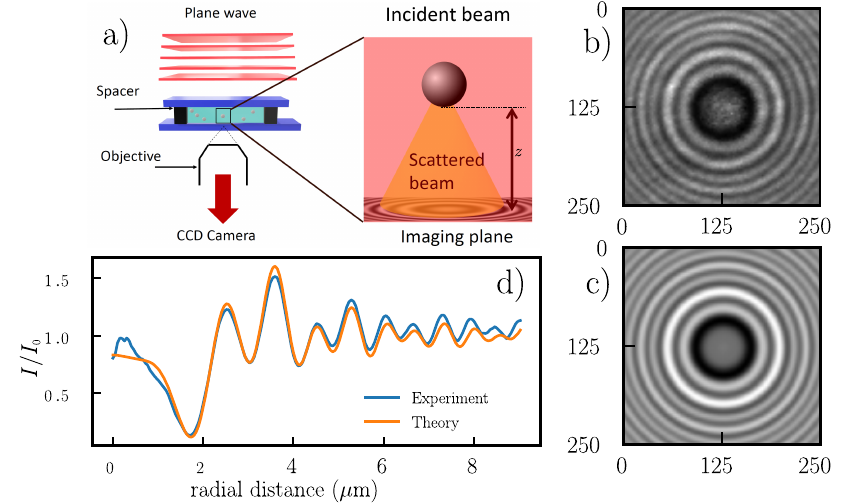}
\caption{a) Schematic of the experimental setup. A laser plane wave of intensity $I_0$ illuminates the chamber containing a dilute suspension of microspheres in water. The light scattered by a particle interferes with the incident beam onto the focal plane of an objective lens, that magnifies the interference pattern and relays it to a camera. b) Typical experimental interference pattern produced by one particle. c) Corresponding best-fit Lorenz-Mie interference pattern~\cite{f_bohren_absorption_1998,mishchenko_scattering_2002,lee_characterizing_2007,vanOostrum2011}, providing a distance $z = 11.24 \pm 0.2 ~
\mathrm{\mu m}$ to the wall, as well as the radius $a = 1.518 \pm 0.006 ~ \mathrm{\mu m}$ and refractive index $n = 1.584 \pm 0.006$ of the particle. d) Angular averages of the intensities $I$ (normalized by $I_0$) from the experimental and theoretical interference patterns, as functions of the radial distance to the $z$ axis.}
\label{Fig2}
\end{figure}

\begin{figure}[t!]
\centering
\includegraphics[scale=1]{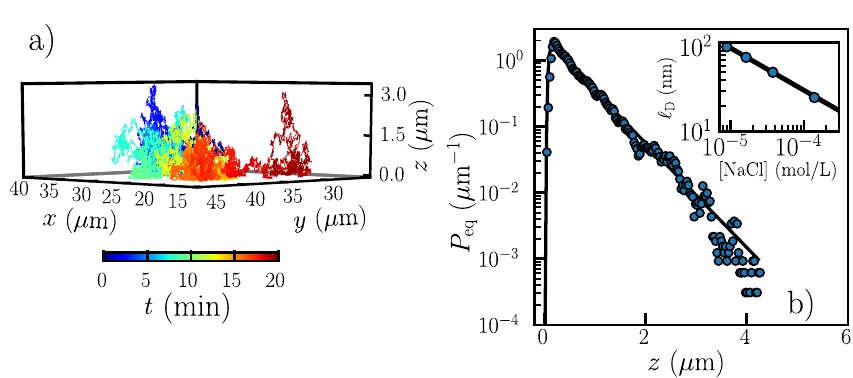}
\caption{a) Typical measured tridimensional trajectory $\bold{r}(t)=[x(t),y(t),z(t)]$ of the microparticle near the wall ($z=0$). b) Measured equilibrium probability density function $P_{\textrm{eq}}$ of the distance $z$ between the particle and the wall. The solid line represents the best fit to the normalized Gibbs-Boltzmann distribution in position, using the total potential energy $U(z)$ of Eq.~(\ref{Eq:PDF}), with $B = 4.8$, $\ell_\mathrm{D} = 21 ~ \mathrm{nm}$, and $\ell_\mathrm{B} = 530~ \mathrm{nm}$. The inset shows the measured Debye length $\ell_\mathrm{D}$ as a function of salt concentration [NaCl]. The solid line is the expected Debye relation $\ell_\mathrm{D}=0.304/\sqrt{\textrm{[Nacl]}}$, for a single monovalent salt in water at room temperature.}
\label{Fig3}
\end{figure}

The experimental setup is schematized in Fig.~\ref{Fig2}a). A sample consists of a parallelepipedic chamber (1.5 cm $\times$ 1.5 cm $\times$ 150 $\mu$m), made from two glass covers, a parafilm spacer, and sealed with vacuum grease, containing a dilute suspension of spherical polystyrene beads (Sigma Aldrich) with nominal radii $a = 1.5\pm0.035~\mathrm{\mu m}$, at room temperature $T$, in distilled water (type 1, MilliQ device) of viscosity $\eta=1$~mPa.s. The sample is illuminated by a collimated laser beam with a $532 ~ \mathrm{\mu m} $ wavelength. The light scattered by one colloidal particle at a given time $t$ interferes with the incident beam. An oil-immersion objective lens (x60 magnification, $1.30$ numerical aperture) collects the resulting instantaneous interference pattern, and relays it to a camera with a 51.6~$\mathrm{nm}$/pixel resolution (see Fig.~\ref{Fig2}b)). The exposure time for each frame is fixed to 3~ms to avoid motion-induced blurring of the image. The angular average of the intensity profile from each time frame is then fitted (see Figs.~\ref{Fig2}c,d)) to the Lorenz-Mie scattering function~\cite{f_bohren_absorption_1998,mishchenko_scattering_2002,lee_characterizing_2007,vanOostrum2011}, which provides the particle radius $a$, its refractive index $n$, and its instantaneous tridimensional position $\bold{r}=(x,y,z)$. To reduce the uncertainty on the position measurement, we first calibrate $a = 1.518 \pm 0.006 ~ \mathrm{\mu m}$ and $n = 1.584 \pm 0.006$ separately from the first $10^5$ time frames. The obtained refractive index is consistent with the one reported in~\cite{matse_test_2017}. Then, for each subsequent time frame, the only remaining fitted quantity is $\mathbf{r}$, which allows us to reconstruct the trajectory $\mathbf{r}(t)$ with a nanometric spatial resolution, as shown in Fig.~\ref{Fig3}a). 

Using the trajectory of the particle, one can then construct the equilibrium probability density function $P_{\textrm{eq}}(\bold{r})$ of the position of the particle. We find that it does not depend on $x$ and $y$, but only on the distance $z$ between the particle and the wall. As seen in Fig.~\ref{Fig3}b), an exponential tail is observed at large distance, which is identified to the sedimentation contribution in Perrin's experiment~\cite{perrin_les_2014}, but here with the probability density function of a single particle instead of the concentration field. In contrast, near the wall, we observe an abrupt depletion, indicating a repulsive electrostatic contribution. Indeed, when immersed in water, both the glass substrate and the polystyrene bead are negatively charged. All together, the total potential energy $U(z)$ thus reads:
\begin{equation}
\frac{U(z)}{k_\mathrm{B}T} =  \left\{
\begin{array}{l}
 \displaystyle B\,\textrm{e}^{-\frac{z}{\ell_\mathrm{D}}} + \frac{z}{\ell_\mathrm{B}}\ ,\quad \text{ for } z>0 \\
+\infty\ ,\quad  \text{ for } z\leq 0
\end{array}
\right. \ ,
\label{Eq:PDF}
\end{equation}
where $k_\mathrm{B}$ is the Boltzmann constant, $B$ is a dimensionless number related to the surface electrostatic potentials of the particle and the wall~\cite{prieve_measurement_1999}, $\ell_\mathrm{D}$ is the Debye length, $\ell_\mathrm{B} = {k_\mathrm{B}T/(g\Delta m)}$ is the Boltzmann length, $g$ is the gravitational acceleration, and $\Delta m$ is the (positive) buoyant mass of the particle. From this total potential energy, one can then construct the Gibbs-Boltzmann distribution $P_{\textrm{eq}}(z) = A\exp[-U(z)/(k_\mathrm{B}T)]$ in position, where $A$ is a normalization constant, that fits the data very well, as shown in Fig.~\ref{Fig3}b). Moreover, as shown in the inset of Fig.~\ref{Fig3}b), we verified that we recover the Debye relation $\ell_{\mathrm{D}}=0.304/\sqrt{\textrm{[Nacl]}}$, with $\ell_{\mathrm{D}}$ in nm, and where [NaCl] is the concentration of salt in mol/L, with a prefactor corresponding to a single monovalent salt in water at room temperature~\cite{Israelachvili2011}. Besides, we have verified (not shown) that the dimensionless parameter $B = 4.8$ related to surface charges is constant in the studied salt-concentration range, thus excluding any nonlinear effect~\cite{Wang2011,Oberholzer1997} in our case. 

We now turn to dynamical aspects, by considering the mean-squared displacement (MSD). For the three spatial directions, indexed by $i=x$, $y$, and $z$, corresponding to the coordinates $r_x=x$, $r_y=y$, and $r_z=z$, of the position $\bold{r}$, and for a given time increment $\Delta t$, the MSD is defined as:
\begin{equation}
\langle\Delta r_i(t)^2 \rangle_t = \langle[r_i(t+\Delta t) - r_i(t)]^2\rangle_t\ ,
\label{MSDdef}
\end{equation}
where the average $\langle\rangle_t$ is performed over time $t$. For a free Brownian motion in the bulk, and in the absence of other forces than the dissipative and random ones, the MSD is linear in time, \textit{i.e.} $\langle\Delta r_i(t)^2 \rangle_t = 2 D_0 \Delta t$, where $D_0=k_\mathrm{B}T/(6\pi\eta a)$ is the bulk diffusion coefficient given by the Stokes-Einstein relation~\cite{einstein_uber_1905}, and $\eta$ is the liquid viscosity. Further including sedimentation restricts the validity of the previous result along $z$ to short times only, \textit{i.e.} for $\Delta t\ll\ell_{\textrm{B}}^2/D_0$ such that the vertical diffusion is not yet affected by the gravitational drift.

The presence of a rigid wall at $z=0$ adds a repulsive electrostatic force along $z$. It also decreases the mobilities nearby through hydrodynamic interactions, leading to effective viscosities $\eta_\parallel(z)=\eta_x(z)=\eta_y(z)$, and $\eta_z(z)$. The latter are~\cite{brenner_slow_1961}:
\begin{equation}
\label{etapara}
\eta_\parallel= \frac{\eta}{1 - \frac{9}{16} \xi + \frac{1}{8} \xi^3 - \frac{45}{256} \xi^4 - \frac{1}{16} \xi ^5}\ ,
\end{equation}
where $\xi = a/(z+a)$, and:
\begin{equation}
\label{etaperp}
\eta_z = \eta\, \frac{6z^2 + 9az + 2a^2}{6z^2 + 2az}         \ ,
\end{equation}
which is Pad\'e-approximated within 1\% accuracy~\cite{bevan_hindered_2000}. 

\begin{figure}[t!]
\centering
\includegraphics{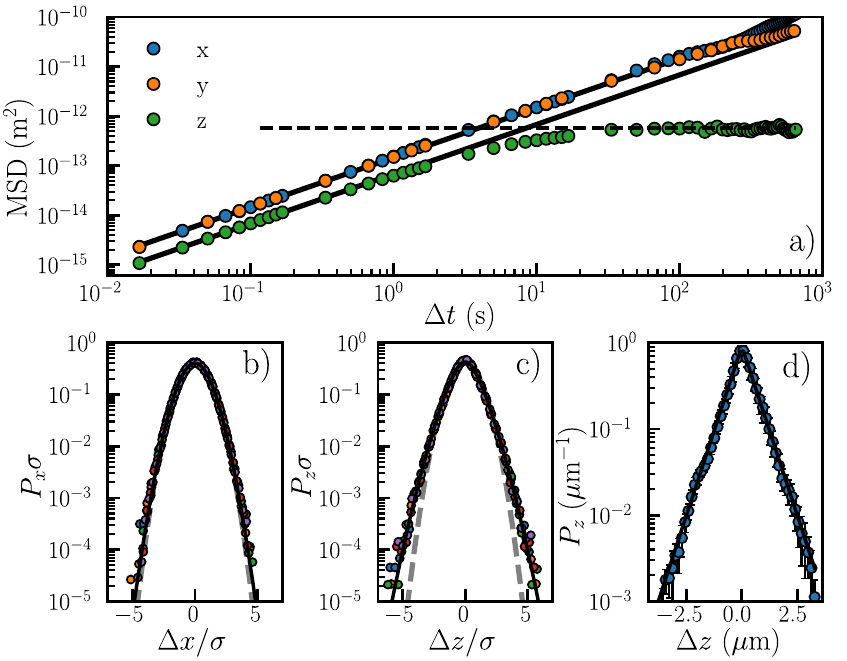}
\caption{a) Measured mean-squared displacements (MSD, see Eq.~(\ref{MSDdef})) as functions of the time increment $\Delta t$, for the three spatial directions, $x$, $y$, and $z$. The solid lines are best fits to Eq.~(\ref{averagediff}), using Eqs.~(\ref{Eq:PDF}),~(\ref{etapara}),~and~(\ref{etaperp}), with $B = 4.8$, $\ell_\mathrm{D} = 21 ~ \mathrm{nm}$, and $\ell_\mathrm{B} = 530~\mathrm{nm}$,
providing the average diffusion coefficients $\langle{D_\parallel}\rangle= \langle D_x\rangle=\langle D_y \rangle =0.52\,D_0$ and $\langle D_z \rangle =0.24\, D_0$. The dashed line is the best fit to Eq.~(\ref{Eq:plateau}), using Eq.~(\ref{Eq:PDF}), with $B = 4.8$, $\ell_\mathrm{D} = 21 ~ \mathrm{nm}$, and $\ell_\mathrm{B} = 530~\mathrm{nm}$. b,c) Normalized probability density functions $P_i\,\sigma$ of the normalized displacements $\Delta x/\sigma$ and $\Delta z/\sigma$, at short times, with $\sigma^2$ the corresponding MSD (see panel a)), for different time increments $\Delta t$ ranging from 0.0167~s to 0.083~s, as indicated with different colors. The solid lines are the best fits to Eq.~(\ref{Eq:PDzshort}), using Eqs.~(\ref{Eq:PDF}),~(\ref{etapara}),~and~(\ref{etaperp}), with $B = 4.8$, $\ell_\mathrm{D} = 21 ~ \mathrm{nm}$, and $\ell_\mathrm{B} = 530~\mathrm{nm}$. For comparison, the grey dashed lines are normalized Gaussian distributions, with zero means and unit variances. d) Probability density function $P_z$ of the displacement $\Delta z$, at long times, averaged over several values of $\Delta t$ ranging between 25 and 30~s. The solid line is the best fit to Eq.~(\ref{auxiliary}), using Eq.~(\ref{Eq:PDF}), with $B = 4.8$, $\ell_\mathrm{D} = 21 ~ \mathrm{nm}$, and $\ell_\mathrm{B} = 530~\mathrm{nm}$.}
\label{Fig4}
\end{figure}

Interestingly, despite the previous modifications, the temporal linearity of the MSD is not altered by the presence of the wall~\cite{chubynsky_diffusing_2014,prieve_measurement_1999} for $x$ and $y$, as well as at short times for $z$. In such cases, the MSD reads:
\begin{equation}
\langle\Delta r_i(t)^2 \rangle_t = 2 \langle D_i \rangle \Delta t\ ,
\label{averagediff}
\end{equation}
where for each spatial direction we introduced the local diffusion coefficient $D_i(z)=D_0\eta /\eta _i(z)$, and its average $\langle D_i(z) \rangle = \int_0^{\infty} \textrm{d}z\, D_i(z)P_{\textrm{eq}}(z)$ against the Gibbs-Boltzmann distribution in position. As shown in Fig.~\ref{Fig4}a), the MSD measured along $x$ or $y$ is indeed linear in time. By fitting to Eq.~(\ref{averagediff}), using Eqs.~(\ref{Eq:PDF})~and~(\ref{etapara}), we extract an average transverse diffusion coefficient $\langle D_\parallel \rangle= \langle D_x\rangle=\langle D_y \rangle= 0.52\, D_0$. In contrast, along $z$, we identify two different regimes: one at short times, where the MSD is still linear in time, with a similarly-obtained best-fit value of $\langle D_z \rangle= 0.24\, D_0$; and one at long times, where the MSD saturates to a plateau. This latter behaviour indicates that the equilibrium regime has been reached, with the particle having essentially explored all the relevant positions given by the Gibbs-Boltzmann distribution.

Having focused on the MSD, \textit{i.e.} on the second moment only, we now turn to the full probability density function $P_i$ of the displacement $\Delta r_i$. Since, the diffusion coefficient $D_i(z)$ varies as a result of the variation of $z$ along the particle trajectory, $P_i$ exhibits a non-Gaussian behavior, as seen in Figs.~\ref{Fig4}b,c,d). We stress that we even resolve the onset of a non-Gaussian behaviour in $P_x$, by zooming on the large-$\lvert\Delta x\rvert$ wings (not shown). At short times, $P_i$ can be modelled by the averaged diffusion Green's function~\cite{matse_test_2017,hapca_anomalous_2009}:
\begin{equation}
P_i(\Delta r_i) = \int ^\infty _0 \mathrm{d}z\, P_{\textrm{eq}}(z) \frac{1}{\sqrt{4 \pi D_i(z) \Delta t}} \textrm{e}^{-\frac{\Delta r_i^2}{4 D_i(z) \Delta t}     }\ ,
\label{Eq:PDzshort}
\end{equation}
against the Gibbs-Boltzmann distribution. As shown in Figs.~\ref{Fig4}b,c), Eq.~(\ref{Eq:PDzshort}) captures the early data very well. At long times, Eq.~(\ref{Eq:PDzshort}) remains valid only for $P_x$ and $P_y$. Nevertheless, the equilibrium regime being reached, $P_z$ can eventually be written as:
\begin{equation}
\lim_{\Delta t\rightarrow\infty}P_z(\Delta z) = \int_0^{\infty}\textrm{d}z\,P_{\textrm{eq}}(z+\Delta z)P_{\textrm{eq}}(z)\ ,
\label{auxiliary}
\end{equation}
which contains in particular the second moment:
\begin{equation}
\lim_{\Delta t\rightarrow\infty}\langle\Delta z ^2\rangle = \int _{- \infty} ^{+ \infty}\textrm{d}\Delta z\, \Delta z^2\int_0^{\infty}\textrm{d}z\, P_{\textrm{eq}}(z+\Delta z)P_{\textrm{eq}}(z) \ .   
\label{Eq:plateau}
\end{equation}
As shown in Fig.~\ref{Fig4}d), Eq.~(\ref{Eq:plateau}) captures the long-term data along $z$ very well.

We now wish to go beyond the previous average $\langle D_i\rangle$ of Eq.~(\ref{averagediff}), and resolve the local diffusion coefficient $D_i(z)$. To measure local viscosities from experimental trajectories, a binning method is generally employed~\cite{friedrich_approaching_2011}. Although this technique is well suited for drift measurements, it suffers from a lack of convergence and precision when second moments or local diffusion coefficients have to be extracted~\cite{frishman_learning_2020}. In particular, the binning method did not allow us to measure specifically the local diffusion coefficient in the key interfacial region corresponding to $z<100$~nm. Besides, Frishman and Ronceray have recently developed a robust numerical method using stochastic force inference, in order to evaluate spatially-varying force fields and diffusion coefficients, from the information contained within the trajectories~\cite{frishman_learning_2020}. In practice, this is done by projecting the diffusion tensor onto a finite set of basis functions. We implemented this method, using fourth-order polynomials in our case. It allowed us to infer the local diffusion coefficients $D_i(z)$, down to $z=10$~nm, as shown in Fig.~\ref{Fig1}a). The results are in excellent agreement with the theoretical predictions, $D_{\parallel}(z)=D_0\eta/\eta_{\parallel}(z)$ and $D_z(z)=D_0\eta/\eta_z(z)$, using the effective viscosities of Eqs.~(\ref{etapara})~and~(\ref{etaperp}), thus validating the method.

So far, through Figs.~\ref{Fig1}a),~\ref{Fig3}b)~and~\ref{Fig4}, we have successively presented the various measured statistical quantities of interest, as well as their fits to corresponding theoretical models. Therein, we have essentially three free physical parameters, $B$, $\ell_\mathrm{B}$, $\ell_\mathrm{D}$, describing the particle and its environment, as well as the \textit{a priori} undetermined location of the $z=0$ origin. These four parameters are actually redundant among the various theoretical models. Therefore, in order to measure them accurately, we in fact perform all the fits simultaneously,
using a Broyden-Fletcher-Goldfarb-Shanno (BFGS) algorithm that is well suited for unconstrained
nonlinear optimization~\cite{dai_convergence_2002}. To do so, we construct a global minimizer:
\begin{equation}
\chi ^ 2 = \sum _{n=1} ^{N} \chi_n ^ 2\ ,
\end{equation}
where we introduce the minimizer $\chi _n ^2$ of each set $n$ among the $N$ sets of data, defined as:
\begin{equation}
\chi _n ^2 = \sum _{i=1} ^{M_n} \frac{[y_{ni} - f_n(x_{ni}, \mathbf{b})]^2 }{f_n(x_{ni}\ , \mathbf{b})^2}\ ,
\end{equation}
with $\{x_{ni},y_{ni}\}$ the experimental data of set $n$, $M_n$ the number of experimental data points for set $n$, $f_n$ the model for set $n$, and $\mathbf{b}=(b_1,b_2,...,b_p)$ the $p$ free parameters. In our case, $p=4$, and $\{x_{ni},y_{ni}\}$ represent all the experimental data shown in Figs.~\ref{Fig1}a),~\ref{Fig3}b)~and~\ref{Fig4}. 

Due to strong dependence of the normal diffusion coefficient $D_z$ with $z$, it is possible to find the wall position with a 10-nm resolution, thus overcoming a drawback of the Lorenz-Mie technique which only provides the axial distance relative to the focus of the objective lens. Besides, the three physical parameters globally extracted from the multifitting procedure are: $B = 4.8 \pm 0.6$, $\ell_\mathrm{D} = 21 \pm 1~ \mathrm{nm} $, and $\ell_\mathrm{B} = 530 \pm 2~ \mathrm{nm}$. Using the particle radius $a = 1.518 \pm 0.006 ~ \mathrm{\mu m}$ calibrated from the preliminary fits of the interference patterns to the Lorenz-Mie scattering function (see Figs.~\ref{Fig2}c,d)), and the $1050 ~ \mathrm{kg.m^{-3}}$ tabulated bulk density of polystyrene, we would have expected $\ell_\mathrm{B}=559 ~ \mathrm{nm}$ instead, which corresponds to less than $2\,\%$ error, and might be attributed to nanometric offsets, such as \textit{e.g.} the particle and/or wall rugosities. 

Finally, we investigate the total conservative force $F_z(z)$ acting on the particle along $z$. By averaging the overdamped Langevin equation over a fine-enough $z$-binning grid and short enough time interval $\Delta t$, one gets in the It$\bar{\textrm{o}}$ convention (corresponding to our definition of $\Delta z$):
\begin{equation}
F_z (z) = 6 \pi \eta_z (z ) a \frac{\langle\Delta z\rangle}{\Delta t} - k_\mathrm{B}T \frac{D_z'(z)}{D_z(z)} \ ,
\label{stokes}
\end{equation}
where the last term corresponds to the additional contribution due to the non-trivial integration of the multiplicative noise~\cite{volpe_influence_2010,mannella_comment_2011,volpe_volpe_2011,Mannella2012}, with the prime denoting the derivative with respect to $z$. From the averaged measured vertical drifts $\langle\Delta z\rangle$, and invoking Eq.~(\ref{etaperp}), one can reconstruct $F_z(z)$ from Eq.~(\ref{stokes}), as shown in Fig.~\ref{Fig1}b). We stress that the statistical error on the force measurement is comparable to the thermal-noise limit~\cite{liu_subfemtonewton_2016}: 
\begin{equation}
\label{tnl}
	\Delta F=\sqrt{24\pi k_{\mathrm{B}}T \eta_z(z) a/ \tau_{\textrm{box}}(z)}\ ,
\end{equation}
 where $\tau_{\textrm{box}}(z)$ is the total time spent by the particle in the corresponding box of the $z$-binning grid. To corroborate these measurements, we invoke Eq.~(\ref{Eq:PDF}) and express the total conservative force $F_z(z)=-U'(z)$ acting on the particle along $z$:
\begin{equation}
\displaystyle F_z(z) =  k_{\mathrm{B}}T\left(\frac{B}{\ell_\mathrm{D}} \textrm{e}^{-\frac{z}{\ell_\mathrm{D}}} - \frac{1}{\ell_\mathrm{B}}\right)\ .
\label{Eq:Force}
\end{equation} 
Using the physical parameters extracted from the above multifitting procedure, we plot Eq.~(\ref{Eq:Force}) in Fig.~\ref{Fig1}b). The agreement with the data is excellent, thus showing the robustness of the force measurement. In particular, we can measure forces down to a distance of $40$~nm from the surface. Besides, far from the wall, we are able to resolve the actual buoyant weight $F_{\textrm{g}} =- 7  \pm 4 ~ \mathrm{fN}$ of the particle. This demonstrates that we reach the femtoNewton resolution, and that this resolution is solely limited by thermal noise.

To conclude, we have successfully built a multi-scale statistical analysis for the problem of freely diffusing individual colloids near a rigid wall. Combining the equilibrium distribution in position, time-dependent non-Gaussian statistics for the spatial displacements, a novel method to infer local diffusion coefficients, and a multifitting procedure, allowed us to reduce drastically the measurement uncertainties and reach the nanoscale and thermal-noise-limited femtoNewton spatial and force resolutions, respectively. The ability to measure tiny surface forces, locally, and at equilibrium, as well the possible extension of the method to non-conservative forces and out-of-equilibrium settings~\cite{Amarouchene2019,Mangeat2019}, opens fascinating perspectives for nanophysics and biophysics.

We thank Elodie Millan, Louis Bellando de Castro, Julien Burgin, Bernard Tr\'egon, Abdelhamid Maali, David Dean and Mathias Perrin for interesting discussions. We acknowledge funding from the Bordeaux IdEx program - LAPHIA (ANR-10IDEX-03-02), Arts et Science (Sonotact 2017-2018) and R\'egion Nouvelle Aquitaine (2018-1R50304).

\bibliographystyle{unsrt}
\bibliography{Lavaud2020}
\end{document}